\documentclass[twocolumn,showpacs,preprintnumbers,amsmath,amssymb]{revtex4}
\usepackage{stmaryrd}%
\usepackage{graphicx}
\usepackage{bm}
\usepackage{graphicx,amsmath,mathrsfs}
\usepackage{amsmath}
\usepackage{amsfonts}
\usepackage{amssymb}%
\providecommand{\U}[1]{\protect\rule{.1in}{.1in}}
\input epsf.sty

\begin{document}

\title{A Separable Pairing Force for Relativistic Quasiparticle Random Phase Approximation}
\author{Yuan Tian$^{1,2}$, Zhong-yu Ma$^{1,3}$, Peter Ring$^{2}$}
\affiliation{(1)~China Institute of Atomic Energy, Beijing 102413,
P.R.of China}
\affiliation{(2)~Physikdepartment, Technische
Universit\"at M\"unchen, D-85748, Garching, Germany}
\affiliation{(3)~Centre of Theoretical Nuclear Physics, National
Laboratory of Heavy Collision, Lanzhou 730000, P.R.of China}

\begin{abstract}
We have introduced a separable pairing force, which was adjusted to
reproduce the pairing properties of the Gogny force in nuclear
matter. This separable pairing force is able to describe in
relativistic Hartree-Bogoliubov (RHB) calculations the pairing
properties in the ground state of finite nuclei on almost the same
footing as the original Gogny interaction. In this work we
investigate excited states using the Relativistic Quasiparticle
Random Phase Approximation (RQRPA) with the same separable pairing
force. For consistency the Goldstone modes  and the convergence with
various cutoff parameters in this version of RQRPA are studied. The
first excited $2^+$ states for the chain of Sn-isotopes with $Z=50$
and the chain of isotones with $N=82$ isotones are calculated in
RQRPA together with the $3^-$ states of Sn-isotopes. Comparing with
experimental data and with the results of original Gogny force we
find that this simple separable pairing interaction is very
successful in depicting the pairing properties of vibrational
excitations.
\end{abstract}

\maketitle

\section{introduction}

At present Density Functional Theory (DFT) based on the mean-field
concept has been widely used for all kinds of quantum mechanical
many-body systems. In nuclear physics the relativistic mean field
theory based on DFT has been of great success in describing the
properties of many nuclei all over the periodic
table~\cite{LNP641.04}. Conventional DFT with a functional $E[\rho]$
depending only on the single particle density $\rho$ can be applied
in nuclear physics practically only in a few doubly closed shell
nuclei. The vast majority of nuclei and in particular those far away
from the $\beta$ stability line the inclusion of particle-particle
($pp$) correlations is essential for a quantitative description of
structure phenomena. In the framework of DFT pairing correlations are
taken into account in the form of Bogoliubov
theory~\cite{BHR.03,VALR.05} for the ground states and in
Quasiparticle Random Phase Approximation (QRPA) for the excited
states. Although monopole pairing or den\-sity dependent delta
pairing interactions are widely used because of their simplicity, a
cutoff parameter has to be introduced by hand. In order to avoid the
complicated problem of a pairing cutoff the finite range Gogny force
has been applied~\cite{DG.80,GEL.96}. Its parameters have been
adjusted very carefully in a semi-phenomenological way to
characteristic properties of the microscopic effective interactions
and to experimental data~\cite{BGG.84,BGG.91}. Over the years the
relativistic Hartree-Bogoliubov theory (RHB)~\cite{GEL.96} with
finite Gogny pairing force has turned out to be a very successful way
to describe pairing correlations in nuclei. The price we have to pay
for the advantages is much more numerical efforts involved,
especially in calculations of deformed nuclei and in applications for
excited states.

As presented in Refs.~\cite{TMR.09,TM.06}, we have introduced a new
separable form of the pairing force for RHB theory. A similar ansatz
has been used in the pairing channel of non-relativistic Skyrme
calculations in Refs.~\cite{DL.08,LDBM.09}. The parameters of our
separable force are adjusted to reproduce the pairing properties of
the Gogny force in nuclear matter. It preserves translational
invariance and it has finite range.  Applying well known techniques
of Talmi and Moshinsky~\cite{Tal.52,Mos.59,BJM.60,BD.66} this
pairing interaction can be used in relativistic and in
non-relativistic Hartree-Bogoliubov or Hartree-Fock-Bogoliubov
calculation of finite nuclei. It avoids the complicated problem of a
cutoff at large momenta or energies inherent in other zero range
pairing forces. In Ref.~\cite{TMR.09} it has been shown that with
this force the pairing properties of ground states can be well
depicted on almost the same footing as with the original Gogny
pairing interaction.

For excited sates it is important to combine the RPA and the RHB in a
consistent way in order to describe the excitations in unstable
nuclei near drip-line, especially when the pairing correlations play
a crucial role. Recently Ring et al.~\cite{RMG.01,PRN.03} have used
time-dependent relativistic mean field theory to derive the fully
self-consistent relativistic random phase (RRPA) and relativistic
quasiparticle random phase approximation (RQRPA). For the pairing
channel the finite range Gogny force D1S is used. Excited states are
calculated in a consistent framework using the same density
functional. It has been shown in several applications that RQRPA
provides an excellent tool for the description of the multipole
response of stable as well as of unstable and weakly bound nuclei far
from stability. These investigations have been devoted to low-lying
collective excitations~\cite{MWG.02,Ans.05,AR.06}, to giant
resonances~\cite{VWR.00,Pie.00,MGW.01,Pie.01,Pie.02}, to spin-isospin
resonances~\cite{PRN.03,PNV.04}, and to new exotic modes in
stable~\cite{VPR.02} and unstable
nuclei~\cite{VPR.01a,PNVR.05,PVR.05,KPA.07,PVK.07}.

Of course, in the case of spherical nuclei, the calculation of
QRPA-matrix elements of the original Gogny force is possible and
computer codes are available~\cite{PRN.03}. Although the evaluation
of such matrix elements for the new separable force is much faster,
nonetheless the application of this force for QRPA-calculations in
spherical nuclei does not bring an essential advantage. This is,
however, no longer true for QRPA-calculations in deformed
nuclei~\cite{PR.08}. Here one has to do with several thousands
two-quasiparticle configurations and several millions of matrix
elements, in particular in relativistic applications where the Dirac
sea has to be treated properly~\cite{RMG.01}. A separable force is
also of considerable advantage in all cases where the RPA-problem
cannot be solved by diagonalization, as for instance for energy
dependent self energies in the treatment of complex configurations
by particle-vibrational coupling~\cite{LRT.08}. In this case one has
to work at fixed energy and to solve the linear response equations
at fixed energy~\cite{RS.80}. It is well known~\cite{RRE.84} that
the dimension of the coupled linear response equations scales with
the number of separable terms and not with the number of
two-quasiparticle configurations. Therefore a separable force brings
essential advantages in all these cases.

So far, the separable pairing force has been used only in static
applications~\cite{TMR.09} and in this case was very successful. It
is not clear from the beginning, whether one can also reproduce the
dynamic properties of full Gogny pairing in such a simple way,
because, in fact, as shown in Fig.~6 of Ref.~\cite{TMR.09} both
forces are not fully identical. In particular there is the problem
of Goldstone modes connected with translational symmetry. It is well
known that these modes depend in a very delicate way on the
properties of the residual interaction. Only in the case of full
self-consistency these modes decouple fully from the rest of the
spectrum. This is particular important for isoscalar dipole
excitations, where the large strength of the spurious translational
mode can contaminate the low-lying E1-spectrum considerably.
Translational invariance is one of the essential advantages of the
new pairing force as compared to older separable pairing forces such
as monopole, quadrupole, or other multipole pairing forces. However,
the new force is presented as a sum over separable terms and
translational invariance is strictly fulfilled only for an infinite
number of separable terms. As it has been shown in
Ref.~\cite{TMR.09}, in static applications this series converges
quickly and one needs only 8 separable terms to get convergence. It
is not clear, wether this number is large enough for a proper
treatment of the Goldstone modes.

This paper is devoted to an investigation of all these open
questions. The new separable pairing interaction is implemented in
the relativistic QRPA program and details for the calculation of the
new $pp$-matrix elements are presented. In order to test the
numerical implementation of the RQRPA equation with the separable
pairing interaction we study the Goldstone modes and the consistency
of the method. In addition we investigate the question whether the
dynamic properties of pairing correlations in vibrational excitations
can be reproduced with the new pairing force. As it is known the
first $2^+$ excited states in semi-magic nuclei are very sensitive to
the pairing gap. Therefore we investigate the isoscalar quadrupole
excitations in Sn-isotopes and in $N=82$ isotones in the RHB+RQRPA
approach with the new pairing force and compare the first 2$^+$
states with those obtained with the full Gogny pairing force.
Furthermore we calculate $3^-$ excitations in Sn-isotopes and
investigate the sensitivity of the isoscalar octupole states to the
pairing properties.

The paper is arranged as follows. The theoretical formalism of
RHB+RQRPA with the separable form of the pairing interaction is
presented in Sec.~II. The consistency of the method as well as the
Goldstone spurious modes are investigated in Sec.~III. The isoscalar
quadrupole in Sn-isotopes and in $N=82$ as well as the isoscalar
octupole states in Sn-isotopes are calculated in the RHB+RQRPA
approach, which are discussed in Sec.~IV. Finally we shall give a
brief summary in Sec.~V.

\section{theoretical formalism}

We start with the  $^{1}S_{0}$ channel gap equation in symmetric
nuclear matter at various densities,
\begin{equation}
\Delta(k)=-\int_{0}^{\infty}\frac{k^{\prime2}dk^{\prime}}{2\pi^{2}%
}\langle{k}|V_{\rm sep}^{^{1}S_{0}}|k^{\prime}%
\rangle\frac{\Delta(k^{\prime})}{2E(k^{\prime})}~, \label{gapeq}%
\end{equation}
where
\begin{equation}
\langle{k}|V_{\rm sep}^{^{1}S_{0}}|k^{\prime
}\rangle=-Gp(k)p(k^{\prime})~ \label{sp}
\end{equation}
is the separable form of the pairing force introduced in
Ref.~\cite{TMR.09} with a Gaussian ansatz $p(k)=e^{-a^{2}k^{2}}$. A
The two parameters $G$ and $a$ are fitted to the density dependence
of the gap at the Fermi surface $\Delta(k_{F})$ in nuclear matter.
Comparing with the Gogny D1S force~\cite{BGG.91}, we obtain the
parameter set of $G=728$ MeV$\cdot$fm$^{3}$ and $a=0.644$ fm.

The RQRPA is constructed in the canonical single-nucleon basis, where
the wave functions of the RHB model have BCS form (for details see
Ref.~\cite{PRN.03}). In these calculations the same interactions are
used in the RHB calculation for the nuclear ground state and in the
RQRPA equations for the excited states, as well in the particle-hole
($ph$) as in the $pp$-channel. Since the interaction in the
$ph$-channel is identical to earlier calculations~\cite{PRN.03}, we
discuss here only the derivation of the matrix elements of the
separable interaction of Eq.~(\ref{sp}) used in the $pp$-channel of
the RQRPA equation in finite nuclei. First, we transform the
separable force Eq.~(\ref{sp}) from momentum space to coordinate
space and obtain
\begin{equation}
V(\mathbf{r}_{1}^{{}},\mathbf{r}_{2}^{{}},\mathbf{r}_{1}^{\prime}%
,\mathbf{r}_{2}^{\prime})=-~G~\delta({{\mathbf{R}}}-{{\mathbf{R}}}^{\prime
})~P(r)P(r^{\prime})~{\scriptstyle\frac{1}{2}}(1-P^{\sigma})~, \label{vr}%
\end{equation}
where
${{\mathbf{R}}}=\frac{1}{2}({{\mathbf{r}}}_{1}+{{\mathbf{r}}}_{2})$
and ${{\mathbf{r}}}={{\mathbf{r}}}_{1}-{{\mathbf{r}}}_{2}$ are the
center of mass and relative coordinates respectively, and $P(r)$ is
obtained from the Fourier
transform of $p(k)$,
\begin{equation}
P(r)=\frac{1}{(4\pi a^{2})^{3/2}}e^{-\frac{r^{2}}{4a^{2}}}~.%
\label{gauss-r}%
\end{equation}
The term $\delta({\mathbf{R}}-{\mathbf{R}}^{\prime})$ in
Eq.~(\ref{vr}) insures the translational invariance. It also shows
that this force is not completely separable in coordinate space.
However, in the basis of harmonic oscillator functions the matrix
elements of this force can be represented by a sum of separable terms
which converges quickly (for details see Ref.~\cite{TMR.09}).

In the pairing channel we need  the two-particle wave functions
coupled to angular momentum $J$ and the projector
$\frac{1}{2}(1-P^{\sigma })$ restricts us to the quantum numbers of
total spin $S=0$ and total orbital angular momentum $\lambda=J$.
These wave functions are usually expressed in terms of the laboratory
coordinates ${\mathbf{r}}_1$ and ${\mathbf{r}}_2$ of the two
particles, while the separable pairing interaction in Eq.~(\ref{vr})
is expressed in the center of mass coordinate ${{\mathbf{R}}}$ and
the relative coordinate ${{\mathbf{r}}}$ of the pair. Therefore we
transform to the center of mass frame by using the well known
Talmi-Moshinsky brackets~\cite{Tal.52,Mos.59,BJM.60} in the notation
of Baranger~\cite{BD.66}
\begin{equation}\label{xxx3}
|n_{1}l_{1},n_{2}l_{2};\lambda\mu\rangle=\sum_{NLnl}M_{n_{1}l_{1}n_{2}l_{2}%
}^{NLnl}|NL,nl;\lambda\mu\rangle,%
\end{equation}
where%
\begin{equation}
M_{n_{1}l_{1}n_{2}l_{2}}^{NLnl}=\langle NL,nl,\lambda|n_{1}l_{1},n_{2}%
l_{2}~,\lambda\rangle
\end{equation}
are the Talmi-Moshinsky brackets with the selection rule%
\begin{equation}
2N+L+2n+l=2n_{1}+l_{1}+2n_{2}+l_{2}.\label{n-selection}%
\end{equation}
Here we need these brackets only for the case $\lambda=J$. We
therefore can express the two-body function with the quantum numbers
$S=0$ and $\lambda=J$ in terms of center of mass and relative
coordinates by the sum
\begin{align}
|12\rangle_{J} & =\frac{\hat{\jmath_1}\hat{\jmath_2}}{\hat{s}}
\left\{
\begin{array}{ccc}j_2&l_2&\frac{1}{2}\\
l_1&j_1&J\end{array}\right\}\sum_{NL}\sum
_{nl}M_{n_{1}l_{1}n_{2}l_{2}}^{NLnl}\label{xxx4}\\
&  ~~~~~\times
R_{NL}(R,b_{R})R_{nl}(r,b_{r})|\lambda=J\rangle|S=0\rangle~,
\nonumber
\end{align}
where $\hat{j}=\sqrt{2j+1}$ and $s=\frac{1}{2}$ and
$R_{NL}(R,b_{R})$, $R_{nl}(r,b_{r})$ are radial oscillator wave
functions for the center of mass and relative coordinates with the
oscillator parameters $b_{R}=b/\sqrt{2}$ and $b_{r}=b\sqrt{2}$.
Finally we find the pairing matrix elements of the interaction
$
V_{121^{\prime}2^{\prime}}^{J}=\langle n_{1}l_{1}j_{1},n_{2}l_{2}%
j_{2}|V|n_{1^{\prime}}l_{1^{\prime}}j_{1^{\prime}},n_{2^{\prime}}l_{2^{\prime
}}j_{2^{\prime}}\rangle_{J}$
~in Eq.~(\ref{vr}) as a sum over the quantum numbers $N,$\ $L,$
$N^{\prime}$, $L^{\prime}$, $n$, $l$, $n^{\prime}$, and $l^{\prime}$
in Eq.~(\ref{xxx3}). The integration over the center of mass
coordinates $\mathbf{R}$ and $\mathbf{R}^{\prime}$ leads to
$N=N^{\prime}$, $L=L^{\prime}$. Further restrictions occur through
the fact
that the sum contains integrals over the relative coordinates of the form%
\begin{equation}
\int R_{nl}(r,b_{r})Y_{lm}(\hat{r})P(r)d^{3}r.
\end{equation}
They vanish unless $l=0$ and $L=J$. The quantum numbers $n$ and
$n^{\prime}$ are determined by the selection rule
(\ref{n-selection}) and we are left with a single sum of separable
terms
\begin{equation}\label{sep}
V_{12,1'2'}^{ppJ}=G\sum_{N}^{N_0}V_{12}^{NJ}\times V_{1'2'}^{NJ}
\end{equation}
where%
\begin{equation}\label{}
\begin{split}
    V_{12}^{NJ}&=\sqrt{4\pi}\frac{\hat{\jmath}_1\hat{\jmath}_2}{\hat{s}}
    \left\{\begin{array}{ccc}
    j_2&l_2&\frac{1}{2}\\ l_1&j_1&J\end{array}\right\}\\
    &\times M_{n_{1}l_{1}n_{2}l_{2}}^{NJn0}\int_0^\infty
    R_{n0}(r,b_r)P(r)r^2dr~.
\end{split}
\end{equation}

For a Gaussian ansatz of $P(r)$ in Eq. (\ref{gauss-r}) this integral
can be evaluated analytically and we find%
\begin{equation}
\begin{split}
    V_{12}^{NJ}&=\frac{1}{b^{3/2}}\frac{2^{1/4}}{\pi^{3/4}}
    \frac{\hat{\jmath}_1\hat{\jmath}_2}{\hat{s}}
    \left\{\begin{array}{ccc}
    j_2&l_2&\frac{1}{2}\\ l_1&j_1&J\end{array}\right\}\\
    &\times\frac{(1-\alpha^{2}%
    )^{n}\text{ \ \ \ \ }}{(1+\alpha^{2})^{n+3/2}}
    M_{n_{1}l_1n_{2}l_2}^{NJn0}~\frac{\sqrt{(2n+1)!}}{2^{n+1}n!}~,%
\end{split}
\end{equation}
where the parameter $\alpha=a/b$ characterizes the width of the
function $P(r)$ in terms of the oscillator length $b$ and $n$ is
given by the selection rule (\ref{n-selection}):
$n=n_{1}+n_{2}+\frac{1}{2}(l_1+l_2-J)-N$. The results of the RHB +
RQRPA model will depend on the choice of the effective RMF Lagrangian
in the $ph$-channel, as well as on the treatment of pairing
correlations. In this work the NL3 effective
interaction~\cite{LKR.97} is adopted for the RMF Lagrangian. In the
pairing channel we use a separable form in Eq.~(\ref{vr}) adjusted to
the pairing part of the Gogny force D1S and compare the results of
RQRPA calculations with the full Gogny force D1S in the pairing
channel.

\section{verification of consistency}

In the following investigations we solve the RHB+RQRPA equations with
this separable pairing force. The Dirac spinors are expanded in a
spherical oscillator basis with $N_F=18$ major oscillator
shells~\cite{GRT.90}. This leads to a very large number of
two-quasiparticle(2$qp$) pairs and to a huge dimension of the
corresponding RQRPA matrix. In the practical numerical calculations a
cutoff energy has to be adopted and 2$qp$-pairs with an energy larger
than this cutoff are neglected. In relativistic RPA we have two types
of 2$qp$-pairs and therefore two cutoff energies: $E_{Cp}$ is the
maximum value of the 2$qp$-energy for positive energy states, and
$E_{Ca}$ is the maximum absolute value of the 2$qp$-energy for states
with one fully or partially occupied state of positive energy and one
empty negative-energy state in the Dirac sea.

To test the numerical implementations of the RHB + RQRPA equations
with the separable form of the pairing interaction we first check the
Goldstone modes~\cite{TV.62}.  As it is known, the Goldstone modes
(often called {\it spurious excitations}) connected with symmetry
violations in the mean field wave function have zero excitation
energy and decouple from the physical states for RPA or
QRPA-calculations based on a self-consistent mean field solution and
using the interaction derived as the second derivative of the energy
density functional~\cite{RS.80}. The importance of a consistent
treatment of pairing correlations in QRPA calculations has been
demonstrated in the non-relativistic~\cite{Mat.01,Mat.02} and in the
relativistic~\cite{DF.90,PRN.03} framework. The zero-energy Goldstone
modes also provide a rigorous check of the consistency. Two kinds of
spurious states have been investigated: one corresponds to the
violation of particle number in the monopole resonance with the
quantum numbers $J^\pi=0^+$; another is connected with the violation
of translational invariance and the spurious center of mass motion in
the dipole resonance with the quantum numbers $J^\pi=1^-$. There
should be no response to the number operator since it is a conserved
quantity, i.e., the Nambu-Goldstone mode associated with the nucleon
number conservation should have zero excitation energy. It is
observed that the spurious state of the number operator in the
nucleus $^{22}$O disappears when the pairing interaction is treated
consistently in the RHB and RQRPA with our separable pairing force.

\begin{figure}[tbp]
\includegraphics[scale=0.25]{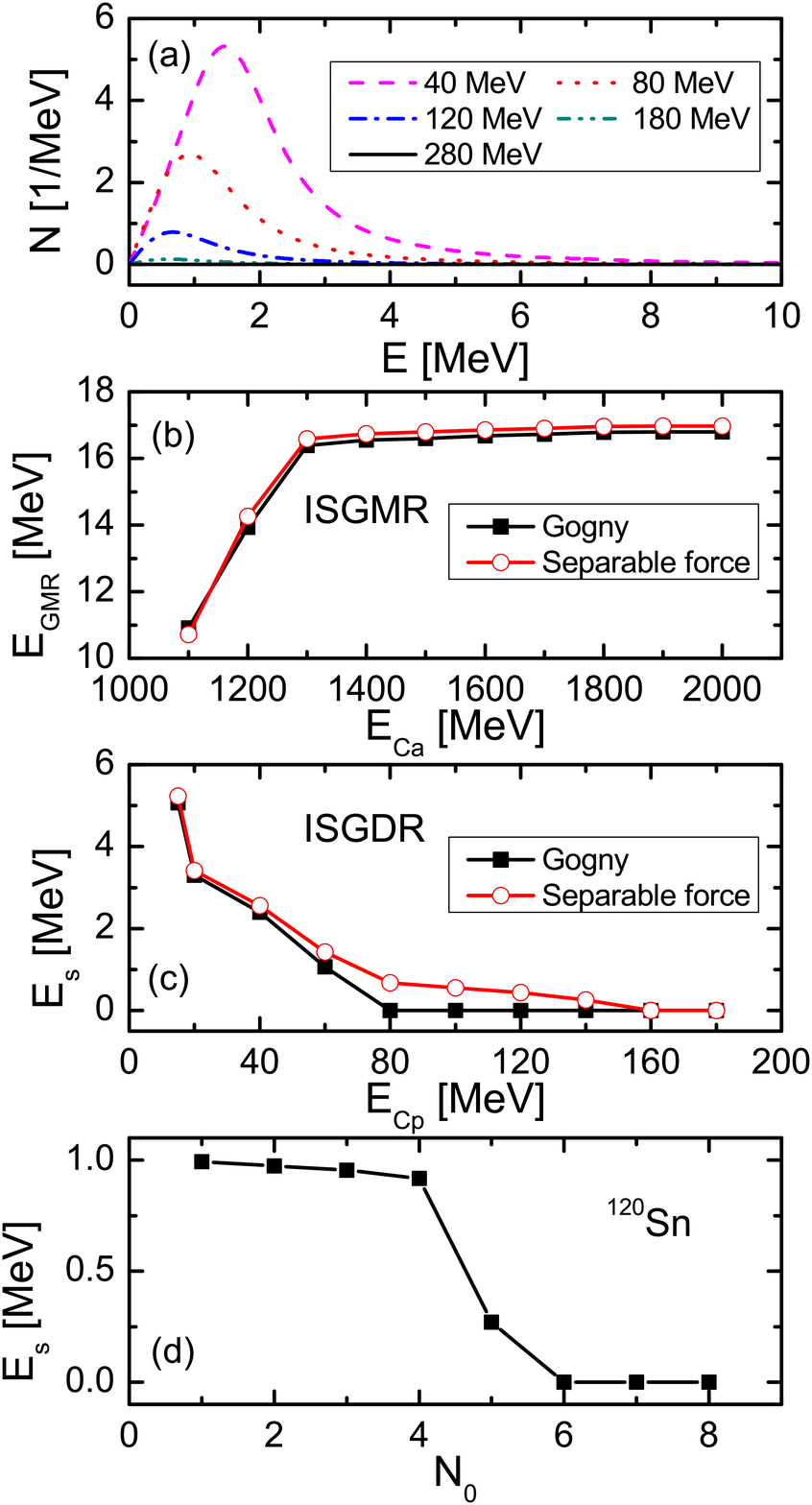}
\caption{(a) The response of the neutron number operator in
$^{120}$Sn for five values of the 2$qp$ cutoff energy parameter
$E_{Cp}$. (b) The excitation energy of the ISGMR in $^{120}$Sn as a
function of the cutoff energy parameter $E_{Ca}$. The red circles and
black squares correspond to the results with the separable pairing
force and the Gogny force, respectively. (c) The position of the
spurious $1^-$ state in $^{120}$Sn as a function of the 2$qp$ cutoff
energy parameter $E_{Cp}$. The notations are the same as in (b). (d)
The position of the spurious $1^-$ state in $^{120}$Sn as a function
of the number $N_0$ of separable terms in Eq. (\ref{sep}).}
\label{fig1}
\end{figure}

For sufficiently large values of the two cutoff parameters $E_{Cp}$
and $E_{Ca}$ the response to the corresponding generator of the
broken symmetry should vanish for all non-vanishing energies. The
investigation of the convergence of the RQRPA results as a function
of these two cutoff parameters provides a very sensitive verification
for the numerical performance of the code. In Fig.~\ref{fig1}(a) we
show how the response to the neutron number operator in $^{120}$Sn
varies with the energy for various values of the cutoff parameter
$E_{Cp}$ in the range from $40-280$ MeV. Here the choice
$E_{Ca}=1800$ MeV includes almost the entire negative-energy Dirac
bound spectrum, which is large enough to yield a convergent result in
the usual RRPA calculations. For $E_{Cp}=180$ MeV the Nambu-Goldstone
$0^+$ mode converges to $\leq 0.1$ MeV. The choice of the cutoff
parameter $E_{Ca}$ has also a pronounced influence on the calculated
isoscalar monopole response. In Fig.~\ref{fig1}(b) we show the peak
energies of the isoscalar giant monopole resonance (ISGMR) in
$^{120}$Sn as a function of $E_{Ca}$. It saturates for
$E_{Ca}\geq1300$ MeV.

In the dipole channel a large configuration space is necessary to
bring the spurious $1^-$ state to zero excitation energy. In
Fig.\ref{fig1} (c) we illustrate the convergence of the energy of the
$1^-$ spurious state in $^{120}$Sn. The excitation energy of the
spurious state is plotted as a function of the energy cutoff
parameter $E_{Cp}$ for a fixed value of $E_{Ca}=1800$ MeV. We find
that the excitation energy goes to zero slightly slower in the case
of the separable pairing force than that with the full Gogny force.
This might be explained by slightly different ranges of the two
pairing forces.

As a consequence we use in the following calculations for the
solution of the self-consistent RHB equations the values $E_{Cp}=180$
MeV and $E_{Ca}=1800$ MeV. This leads to a dimension in the order of
3500 2$qp$-pairs for the RQPRA matrix.

As we see from the Eq.~(\ref{sep}) the separable pairing interaction
is not fully separable in the spherical harmonic oscillator basis. We
have a sum over the quantum number $N$ characterizing the major
shells of the harmonic oscillator in the center of mass coordinate.
In order to study the convergence with the parameter $N_0$, we show
in Fig.~\ref{fig1}(d) the $1^-$ spurious state in $^{120}$Sn as a
function of $N_0$. We find that for nuclei around the line of
$\beta$-stability ($^{120}$Sn), $N_0=5$ is already large enough to
bring the spurious $1^-$ state to zero excitation energy. In the
previous investigation in Ref.~\cite{TMR.09} it was found that a
somewhat larger value of $N_0=8$ is needed to obtain convergence for
the ground state properties.

\section{Results and discussion}

\begin{figure}[tbp]
\includegraphics[scale=0.25]{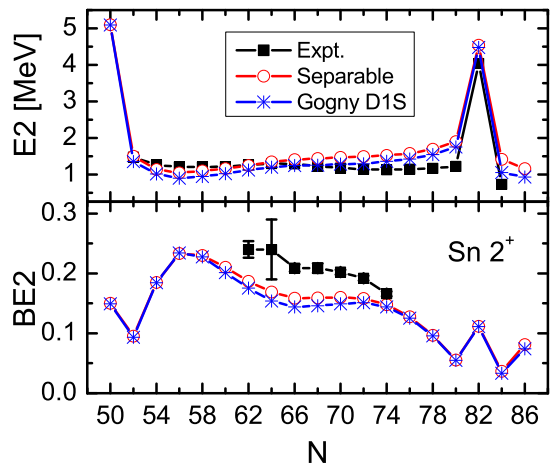}
\caption{Dependence of $E2$ and $B(E2)\uparrow$ on the number of
neutrons for Sn isotopes. The $B(E2)$ values are given in units of
[10$^4$e$^2$fm$^4$]} \label{fig2}
\end{figure}

First we investigate the lowest $2^+$ excitations in Sn-isotopes for
which experimental data are available using the relativistic
parameter set NL3 in the $ph$-channel. In Fig.~\ref{fig2} we plot
the $E2$ excitation energies and the $B(E2)\uparrow$ values for the
chain of Sn-isotopes as a function of the mass number. We compare
the results obtained using the new separable interaction (\ref{sp})
with the calculations by Ansari \cite{Ans.05} using the original
Gogny force D1S in the pairing channel. The results obtained with
the separable pairing interaction are slightly larger than those
calculated with the Gogny force. However, the discrepancy stays very
small within a few percent, and the behavior of the $E2$ and the
$B(E2)$ values along the chain of isotopes is well reproduced. These
small deviations can be understood by the fact that the one term
separable pairing interaction, as we discussed in
Ref.~\cite{TMR.09}, yields in the ground states slightly larger
pairing gaps and therefore stronger pairing fields than those found
with the original Gogny force. Therefore, as expected, also the
effect of the pairing fields in the excited states is slightly
increased in the case of the separable pairing force. This
conclusion is consistent with the pairing properties of the ground
states described in the RHB calculations.

\begin{figure}[tbp]
\includegraphics[scale=0.25]{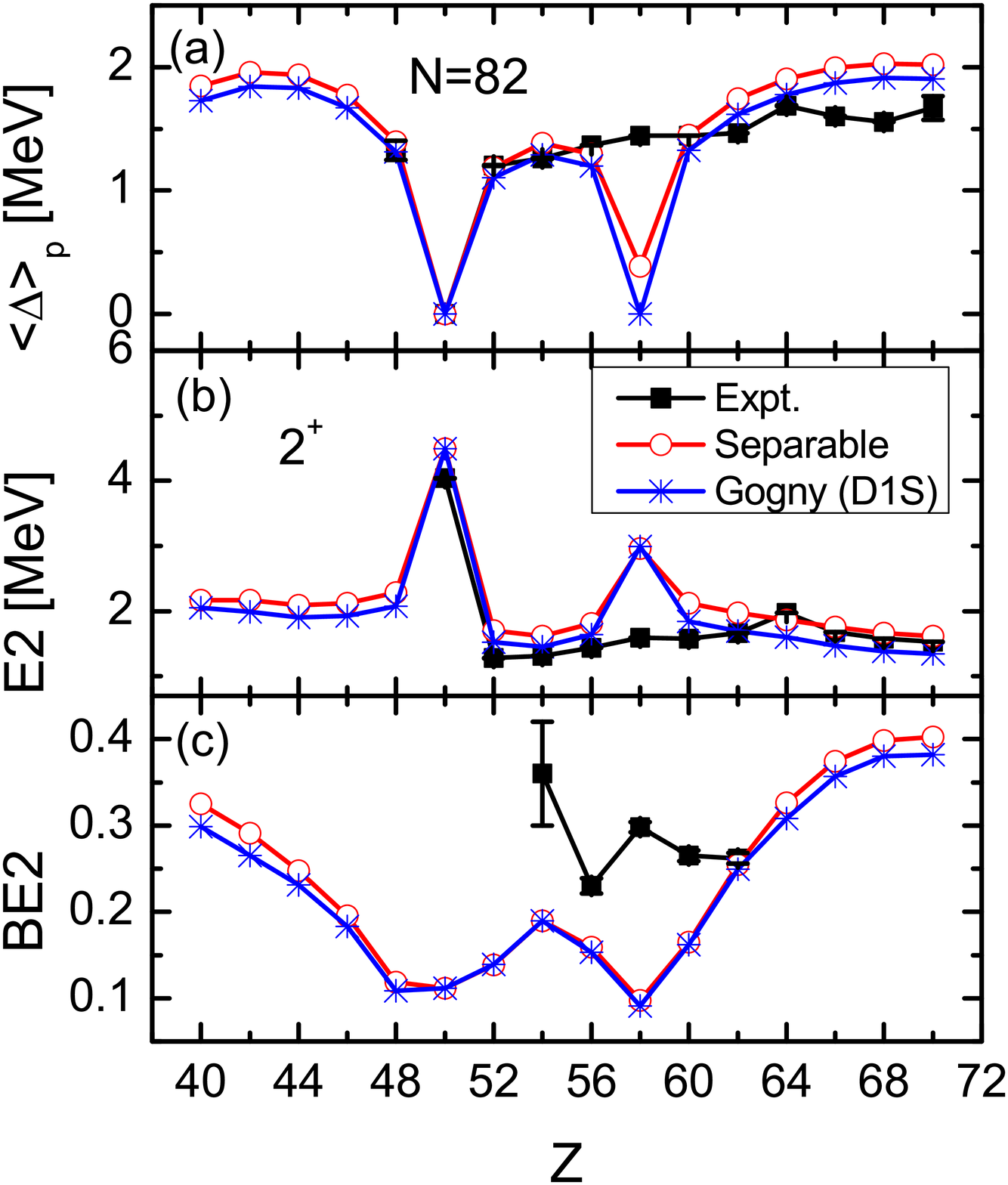}
\caption{(a) The proton average gap of the $N=82$ isotones from
$^{122}_{40}$Zr to $^{152}_{70}$Yb. Dependence of $E2$ (middle) (b)
and $B(E2)$ (bottom) (c) on the number of protons for $N=82$
isotones. The $B(E2)$ values are given in units of
[10$^4$e$^2$fm$^4$].} \label{fig3}
\end{figure}

In the top panel of Fig.~\ref{fig3}(a) we show the average pairing
gaps for protons obtained with RHB calculations for the chain of the
$N=82$ isotones from $^{122}_{40}$Zr to $^{152}_{70}$Yb. We find that
the separable pairing interaction describes the average paring gaps
of finite nuclei on almost the same footing as the Gogny force D1S,
although the pairing gaps calculated with the separable force are
always slightly larger than those with the Gogny force. In the middle
and bottom panels of Fig.~\ref{fig3} we display the $E2$ excitation
energies and the $B(E2)\uparrow$ values for the chain of $N=82$
isotones as functions of the proton number. Again very similar
results are found for the Gogny force and for its separable form in
the pairing channel. Experimental data for the pairing gaps, the
excitation energies of the lowest $2^+$ states, and the corresponding
$B(E2)\uparrow$ values are also plotted in Fig.~\ref{fig3} for the
chain of $N=82$ isotones. The experimental values of the pairing gap
in even-even nuclei are calculated by the odd-even mass difference
with the three-point formula~\cite{AWT.03}. It is shown that the
agreement between theoretical predictions and experimental data is
reasonable, except for the nucleus $^{140}$Ce. Our results for the
RHB-calculations show that the nucleus $^{140}$Ce with the charge
number $Z=58$ has a closed sub-shell for protons: the $\pi 1 g_{7/2}$
state is fully occupied and we observe in the single particle
spectrum for protons a large energy gap at the Fermi surface.
Therefore a very small pairing gap and a large excitation energy of
the lowest $E2$ state were predicted, which is inconsistent with the
experimental data.

\begin{figure}[tbp]
\includegraphics[scale=0.25]{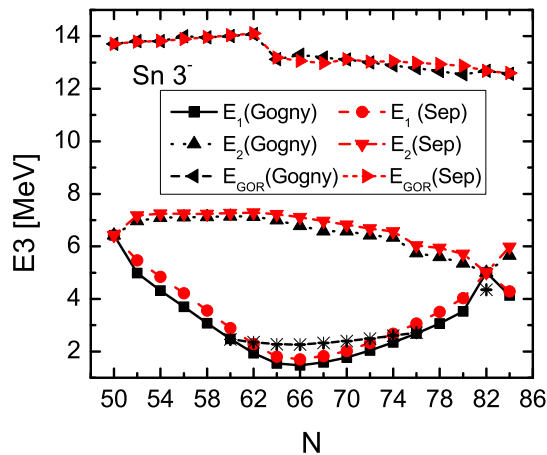}
\caption{Energies of the first ($E_1$) and second ($E_2$) pygmy
states of $E3$ and the peak energies of the Giant Octupole states for
Sn-isotopes as  functions of neutron number. The star line is the
experimental data of first E3 state of Sn-isotope.}\label{fig4}
\end{figure}

We also investigate in the RHB+RQRPA approach excited octupole
states of spherical Sn-isotopes. In Fig.~\ref{fig4} we plot the
first and second excited $J^\pi=3^-$-state and the peak energy of
the giant octupole resonance for the Sn-isotopes as functions of the
neutron number. Strong low-lying $3^-$ states are found also in the
Sn-isotopes, which are consistent with experimental observations. We
can also observe that the results for the $3^-$ states calculated
with the separable pairing force are very close to those obtained
with the Gogny force in Ref.~\cite{AR.06}. The calculations with the
separable pairing force yield slightly larger values of the first
and second excited states than those with the Gogny force as we have
seen it in the case of low-lying quadrupole excitations, while both
of them give almost the same peak energies of the giant octupole
resonances. This is due to the fact that pairing collections have
rather little influence on the giant resonances, but a strong effect
on the low-lying excitations in semi-magic nuclei. In comparison
with the experimental data~\cite{Spear.02}, the RHB+RQRPA
calculations both with Gogny and separable pairing forces can well
describe the first $3^-$ states. This again illustrates that both
the Gogny pairing force and its separable approximation describe the
pairing properties of excited states on almost the same footing.

\section{conclusion}

In summary we have presented first results for RHB+RQRPA calculations
in finite nuclei based on a new separable force in the pairing
channel. This separable force is translational invariant and has
finite range. It contains 2 parameters which are adjusted to
reproduce the pairing gap of the Gogny force in nuclear matter. In
the RHB calculations for finite nuclei the two-body matrix elements
in the $pp$-channel are evaluated using well known techniques of
Talmi and Moshinsky. The separable form of the pairing interaction
can describe the pairing properties of finite nuclei in the ground
states on almost the same footing as its corresponding pairing Gogny
interaction~\cite{TMR.09}. Similar techniques are used for the
evaluation of the two-body matrix elements in the $pp$-channel for
RQRPA calculations. The numerical implementations of the RQRPA with
this separable pairing force are verified by checking for the
separation of Goldstone modes connected with symmetry violations in
the mean field solutions. The numerical convergence of the RQRPA
calculations as a function of various cutoff parameters is shown for
the nucleus $^{120}$Sn. We presented applications to the lowest 2$^+$
states and the corresponding reduced transition rates in Sn isotopes
and $N=82$ isotones. The isoscalar octupole excitations in Sn
isotopes are also investigated. We found excellent agreement of our
results in comparison with those obtained by using the full Gogny
force in the $pp$-channel and with available experimental data.

Therefore we can conclude that this simple separable pairing
interaction can also be applied in future applications of the
RHB+RQRPA approach in nuclei far from stability instead of the
complicated Gogny force. In particular this will allow us to use
realistic, finite range pairing forces also in cases where the
numerical complexity forced us up to now to neglect pairing
correlations completely, as for instance in recent investigations of
magnetic dipole modes based the tilted axis cranking
approach~\cite{PMR.08}, or to restrict us to very simple monopole or
zero range forces in the pairing channel, as for instance in
relativistic QRPA calculations in deformed
nuclei~\cite{PR.08,PKR.09}. There are also many extensions of
relativistic density functional theory beyond mean field, which were
so far only possible with rather simple pairing forces, such as
applications using projection~\cite{YMP.09} onto subspaces with good
symmetries, generator coordinate methods
(GCM)~\cite{NVR.06a,NVR.06b}, or investigation of complex
configurations in the framework of particle-vibrational coupling
(PVC)~\cite{LRT.08,LRT.09}. All these methods require a more
realistic description of pairing correlations in the future.
Investigations in this direction are in progress.

\begin{acknowledgments}
This research has been supported by the National Natural Science
Foundation of China under Grant Nos 10875150, 10775183 and 10535010,
the Major State Basis Research Development of China under contract
number 2007CB815000, the Bundesministerium f\"{u}r Bildung und Forschung
(BMBF), Germany under project 06 MT 246 and the DFG cluster of
excellence \textquotedblleft Origin and Structure of the
Universe\textquotedblright\ (www.universe-cluster.de).
\end{acknowledgments}


\end{document}